# Pathway to oxide photovoltaics via band-structure engineering of SnO

Haowei Peng, Andre Bikowski, Andriy Zakutayev, Stephan Lany
National Renewable Energy Laboratory, Golden, CO 80401

## Abstract

The prospects of scaling current photovoltaic technologies to terawatt levels remain uncertain. All-oxide photovoltaics could open rapidly scalable manufacturing routes, if only oxide materials with suitable electronic and optical properties were developed. A potential candidate material is tin monoxide (SnO), which has exceptional doping and transport properties among oxides, but suffers from a low absorption coefficient due to its strongly indirect band gap. Here, we address this shortcoming of SnO by band-structure engineering through isovalent but heterostructural alloying with divalent cations (Mg, Ca, Sr, Zn). Using first-principles calculations, we show that suitable band gaps and optical properties close to that of direct semiconductors are achievable in such SnO based alloys. Due to the defect tolerant electronic structure of SnO, the dispersive band-structure features and comparatively small effective masses are preserved in the alloys. Initial $Sn_{1-x}Zn_xO$ thin films deposited by sputtering exhibit crystal structure and optical properties in accord with the theoretical predictions, which confirms the feasibility of the alloying approach. Thus, the implications of this work are important not only for terawatt scale photovoltaics, but also for other large-scale energy technologies where defect-tolerant semiconductors with high quality electronic properties are required.

## Introduction

The relatively high factory capital expenditures for silicon based photovoltaics (PV) necessitate enormous financial investments to build the manufacturing capacities required to scale solar energy generation to terawatt levels [1]. After a decade of unprecedented growth, the capital investments peaked in 2011, but strongly consolidated in recent years, raising the question whether the manufacturing capacities will suffice to support projected PV market growth scenarios [2]. As a next generation solar cell concept, all-oxide PV [3] has recently gained attention due to its potential compatibility with cost-effective and highly scalable manufacturing processes, as well as the promise of high chemical stability, module reliability, and, hence, bankable novel PV technologies. However, the development of this field is held back by the lack of oxide materials with suitable electronic properties. An efficient PV absorber material requires an extensive set of both intrinsic and extrinsic properties, including a suitable band gap, high optical absorption within the visible optical spectrum, high carrier mobilities, controllable (ideally bipolar) doping, and the absence of detrimental defect states. Due to these demanding requirements, solar cell efficiencies above 10% have been demonstrated only for a handful of different materials systems, not including any oxides [4].

While electron (*n*-type) conduction is readily available in wide gap oxides like ZnO and $In_2O_3$, good hole (*p*-type) transport properties are much harder to achieve in oxide materials. However, the efficient bipolar transport is very important for successful PV or photo-electrochemical (PEC) applications. Traditionally, most solar energy conversion research on oxides has focused on transition metals oxides with closed electronic shells, such as the $d^{10}$ $Cu_2O$ and related ternary oxides [5, 6], or the $d^0$ $TiO_2$ [7] albeit with lower energy conversion efficiencies. However, as recently shown for the tetrahedral Mn(II) oxides [8, 9, 10], even open-shell systems can exhibit favorable semiconducting hole transport despite the historical notion of being Mott insulators [11]. The reason for good charge transport in materials like $Cu_2O$ or MnO is that *p-d* coupling creates a valence band maximum (VBM) with an anti-bonding character, which can facilitate *p*-type doping [5], defect tolerance [12], and lower hole effective masses [13]. The importance of an anti-bonding VBM for photovoltaics is supported by the fact that the leading oxide solar cell approaches employ the $d^{10}$ oxide $Cu_2O$ as the active absorber layer. Respectable PV performance has been demonstrated [14], even though $Cu_2O$ has highly non-ideal band structure and optical properties [15].

In the low-valent Sn(II) state, the Sn 5*s* orbital is occupied and exhibits an *s-p* coupling with the O 2*p* ligand orbitals [16, 17, 18], not unlike the above mentioned *p-d* interaction. The delocalized character of the 5*s* states leads to an exceptionally strong valence band dispersion and low hole effective masses in SnO, which has enabled its applications in *p*-type thin-film transistor [19]. Interestingly, this atomic orbital interaction mechanism occurs also for Pb(II), Sb(III), and Bi(III) [13, 18]. In fact, it is now recognized as an important factor [20, 21] in the stunning efficiencies achieved within a very short time scale of development in perovskite photovoltaics based on methyl-ammonium lead iodide ($MAPbI_3$) [22]. Due to stability issues in these perovskites [23], it would be highly desirable to transfer this design principle to the realm of oxide photovoltaics. Since, in SnO, the Sn 5*p* like conduction band is also highly dispersive [16] with a small electron effective mass of about 0.4 $m_0$, *p*-type SnO could support long minority lifetimes which is one of the most important metrics for PV application. Furthermore, SnO has attracted attention for its bipolar doping [24], which could in principle enable a homojunction solar cell design.

The single most important reason why SnO has not been considered as a solar energy material is the strongly indirect character of its band gap. As deduced from optical absorption [19] and electronic structure calculations [25], the fundamental band gap $E_g$ is only about 0.7 eV with the valence band maximum (VBM) at the Γ-point and the conduction band minimum (CBM) at the M-point of the tetragonal Brillouin zone. The optical band gap $E_g^O$ is about 2.7 eV and corresponds to direct optical transitions at Γ. For PV absorbers, $E_g$ determines the upper limit of the open circuit voltage $V_{OC}$, and $E_g^O$ limits the short circuit current $J_{SC}$. Thus, the small gap and the high absorption threshold of SnO severely limit the maximum efficiency achievable in a thin film PV device. This situation is comparable to Si ($E_g$ = 1.1 eV, $E_g^O$ = 3.4 eV), where the weak phonon assisted indirect absorption below $E_g^O$ necessitates the use of fairly thick slabs (usually > 100 μm) of high-quality, defect-free material that supports the required photoexcited carrier diffusion length.

Following the interest in SnO for optoelectronic applications, a recent theoretical study [26] has proposed mixed valence $Sn_xO_y$ compounds as solar energy materials, which are formed by stacking of SnO, $Sn_2O_3$ and $Sn_3O_4$ monolayers. However, the precise control of the mixed oxidation states and the layer structure may prove difficult for large-scale applications. Alternatively, alloying through elemental substitution is a well-established industrially relevant means of band gap engineering, which can be utilized to modify the optical properties in a controllable fashion [27]. However, it is of high importance to ensure that favorable band structure properties are maintained upon the elemental substitution. For example, alloying of wide-gap oxides such as $TiO_2$ and ZnO to reduce their band gaps has attracted a lot of interest in photocatalysis [28], but these materials are prone to develop defect states inside the band gap upon alloying [29, 30]. Motivated by the expected defect tolerance of SnO, here we investigate the evolution of the band-structure and optical properties in isovalent $Sn_{1-x}M_xO$ alloys (M = Mg, Ca, Sr, Zn), so to explore the feasibility of SnO based thin film solar cells. The first principles calculations show that the difference between the direct and indirect band gaps is strongly reduced upon such heterostructural alloying, while the band gap region stays clear of defect states. Thin-film deposition and characterization experiments demonstrate the feasibility of synthesis of these materials and provide an initial verification of the theoretically predicted properties. The results of this work lead to the conclusion that SnO-based alloys are promising defect-tolerant semiconductors for photovoltaic solar cells and other energy conversion applications.

## Results and Discussion

### Computational

Figure 1a shows the mixing enthalpies $\Delta H_m$, obtained from density functional theory (DFT) calculations in the generalized gradient approximation (GGA), for the $Sn_{1-x}M_xO$ alloys (M = Mg, Ca, Sr, Zn),

$$\Delta H_m = E(Sn_{1-x}M_xO) - (1-x)E(SnO) - xE(MO), \qquad (1)$$

where $E$ is the calculated total energy. For the pure oxides, we considered the respective ground-state structure, i.e., the litharge structure (space group 129) for SnO, rocksalt for MgO, CaO, and SrO, and wurtzite for ZnO. In the alloys, the cations were exchanged substitutionally in the litharge structure. As found in other heterostructural alloys, [10] the mixing energy increases almost linearly with the composition. Such a linear relation suppresses the spinodal decomposition, thereby providing an opportunity to grow homogeneous alloys beyond the binodal solubility limit via non-equilibrium deposition techniques [10, 31]. The mixing enthalpy decreases from Mg-, Zn-, Ca-, to Sr-alloyed systems (see Fig. 1a), which can be related to the reduction of the ionic size mismatch. In order to compare the ionic radii of the M cations and $Sn^{2+}$ within the same coordination environment, we use the DFT calculated M-O bond-length in the rocksalt structure, i.e., 2.12, 2.35, 2.60, 2.12, and 2.56 Å for Mg, Ca, Sr, Zn, and Sn, respectively. Taking the Shannon radius for $O^{2-}$ (coordination number = 6) of 1.40 Å [32], the cation radii are 0.72, 0.95, 1.20, 0.72, and 1.16, respectively. The ionic radii mismatch and, correspondingly, the mixing energy increase along the sequence Sr-Ca-Zn-Mg. Thus, from these calculations, we expect the highest solubility for Sr, but the moderate magnitude of the mixing energy suggests that even Zn substitution might be feasible in non-equilibrium deposition techniques.

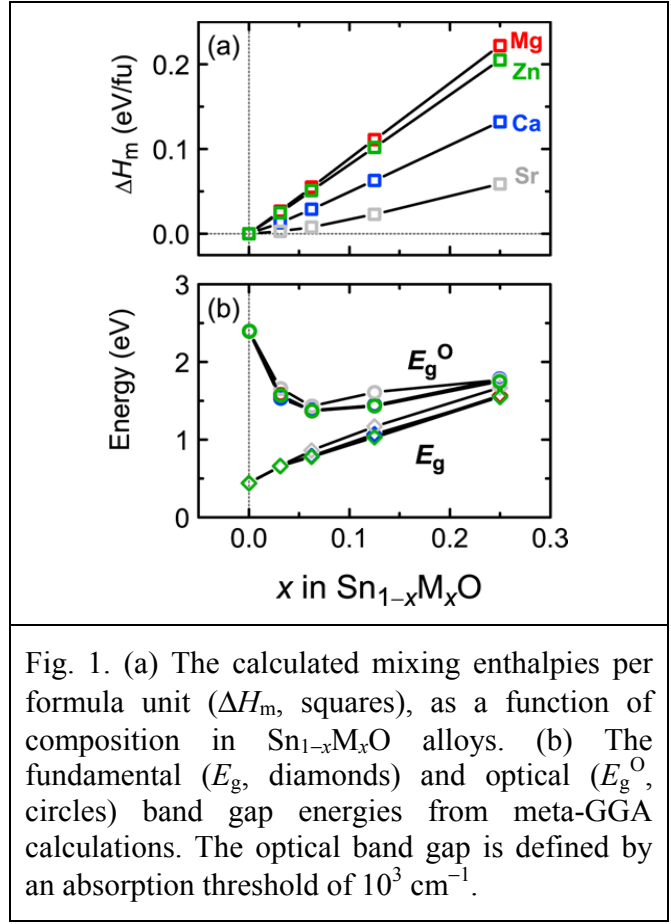

Fig. 1. (a) The calculated mixing enthalpies per formula unit ($\Delta H_m$, squares), as a function of composition in $Sn_{1-x}M_xO$ alloys. (b) The fundamental ($E_g$, diamonds) and optical ($E_g^O$, circles) band gap energies from meta-GGA calculations. The optical band gap is defined by an absorption threshold of $10^3$ cm$^{-1}$.

In order to predict the electronic structure and optical properties of the alloys, we used a meta-GGA functional that largely avoids the band gap error of DFT [33]. Figures 2a and 2b show the total density of states (DOS) and local density of states (LDOS) per atomic site for SnO and $Sn_{1-x}M_xO$ alloys at $x = 0.125$. The Sn-$5s$ and Sn-$5p$ orbital contributions dominate near the VBM and CBM, respectively. The substitution of Sn by the divalent cations increases the fundamental band gap and creates new features inside the valence and conduction bands. In particular, Zn substitution causes a resonance at about 2 eV below the VBM, originating from the Zn-$3d$/O-$2p$ anti-bonding interaction, and Ca causes an increased DOS above the CBM. Importantly, the extrinsic cations do not create gap states or sharp defect-like states near the band edges. Hence, alloying affects the electronic structure and optical properties mainly by reducing the symmetry of the perfect SnO crystal structure and disrupting the original Sn-O and Sn-Sn network (Fig. 2c), which leads to a reduced band dispersion and an opening of the band gap. Accordingly, we should not find a strong chemical dependence of the electronic structure and optical absorption spectrum on the alloying element (Zn, Sr, Ca, Mg) as described below.

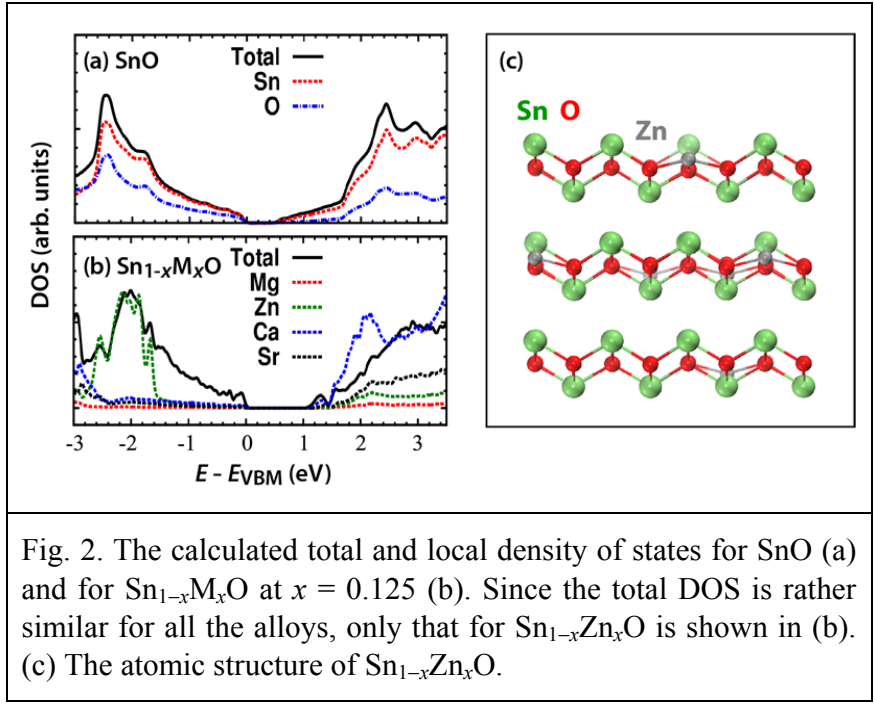

Fig. 2. The calculated total and local density of states for SnO (a) and for $Sn_{1-x}M_xO$ at $x = 0.125$ (b). Since the total DOS is rather similar for all the alloys, only that for $Sn_{1-x}Zn_xO$ is shown in (b). (c) The atomic structure of $Sn_{1-x}Zn_xO$.

The dependence of the fundamental band gap $E_g$ and optical band gap $E_g^O$ on the alloy composition is shown in Fig. 1(b). Here, $E_g^O$ is defined as the energy at which the optical absorption coefficient α exceeds $10^3$ cm$^{-1}$, which is a threshold suitable for thin films with a thickness in the order of a μm. To put this definition into context, $E_g^O$ is about 0.2 eV larger than $E_g$ in GaAs when determined from a similar calculation. As shown in the Fig. 1b, the fundamental band gap increases monotonically with the alloy composition and enters the range of interest for PV applications around $x = 0.1$. In contrast, the optical gap $E_g^O$ shows a non-monotonic behavior, initially decreasing steeply up to $x = 0.05$, and then approaching the fundamental gap, with a $E_g^O - E_g$ difference similar to that in direct gap materials. The trend is very similar for all elemental substitutions, indicating that the increased optical absorption is mostly a result of the alloy disorder, lifting the momentum selection rules that cause the indirectness and large discrepancy between $E_g$ and $E_g^O$ in pure SnO.

It is well known that the degree to which the band structure character is maintained in alloys depends on the strength of the perturbation induced by the elemental substitution [34, 35]. Strong perturbations can effectively destroy the $E(k)$ relationship. The substitution of $Sn^{2+}$ by the divalent cations Mg, Ca, Sr, and Sn is obviously rather strong perturbation, as these ions do not possess the stereoactive lone pair, and usually exhibit a very different coordination environment. For example, Zn assumes a near planar 4-fold coordinated configuration within the O layers (see Fig. 2c). Since the preservation of band dispersion is expected to be of high importance for both majority and minority carrier transport, we calculated the effective band structure for $Sn_{120}Zn_8O_{128}$ and $Sn_{112}Sn_{16}O_{128}$ supercells. Figure 3 shows the spectral function [36],

$$A(\mathbf{k}, E) = \sum_{i,\mathbf{K}} |\langle \psi_{i,\mathbf{K}} | \mathbf{k} \rangle|^2 \delta(\epsilon_i - E) \qquad (2)$$

where **k** denotes the Bloch wave vectors of SnO primitive cell and **K** that of the SQS supercell, $\psi_{i,\mathbf{K}}$ is the eigenstate of the supercell with the eigenvalue $\epsilon_i$, and the spectral weight $|\langle\psi_{i,\mathbf{K}}|\mathbf{k}\rangle|^2$ describes the momentum projection of the supercell eigenstates onto the Brillouin zone of the primitive cell. For comparison, we also plotted the band structure of pristine SnO on top of the effective band structure of the alloys.

We observe that the backbone of the SnO band structure is preserved in the alloys, with a well-defined VBM at Γ and the CBM at M, while the fundamental gap increases with the Zn composition. New states without a clear k-dispersion emerge at energies above the CBM and below the VBM. These states contribute to low-energy direct band-to-band absorption processes, which are not allowed in pristine SnO. Importantly, these new states do not eliminate the dispersive band structure features at the band edges, and we therefore expect that favorable carrier mobilities are preserved. We attribute the absence of alloy-induced gap states and the preservation of band dispersion to the defect-tolerant nature of SnO, arising from bonding–antibonding interactions of Sn and O atomic orbitals [12] (specifically, occupied Sn-5$s$/O-2$p$ interactions in the valence band and unoccupied Sn-5$p$/O-3$s$ interactions in the conduction band).

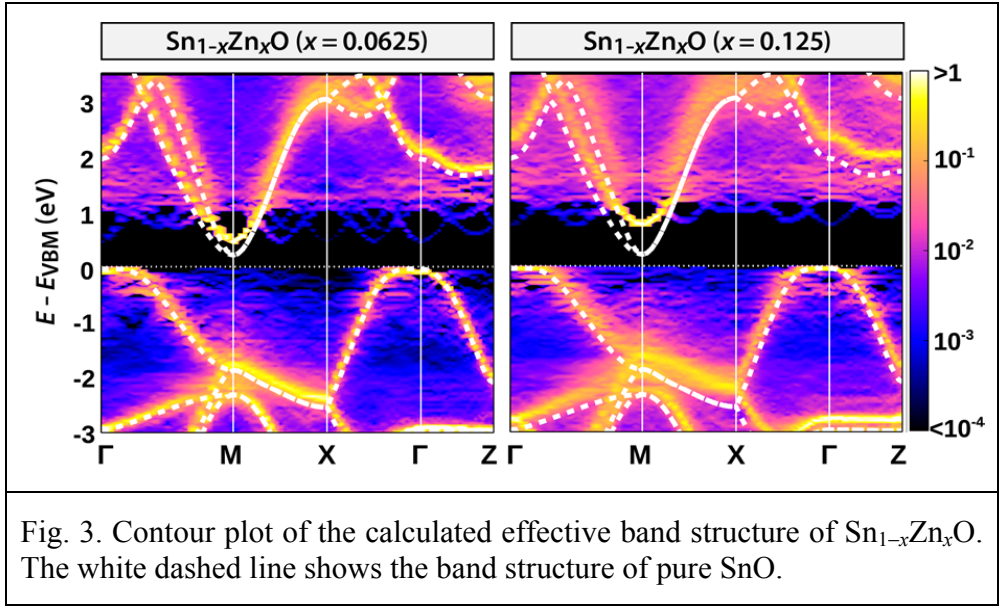

Fig. 3. Contour plot of the calculated effective band structure of Sn$_{1-x}$Zn$_x$O. The white dashed line shows the band structure of pure SnO.

In order to quantify the electronic structure changes near the band edges, we further determined the DOS effective masses [37]. E.g., the hole effective mass is defined as

$$m_h^* = \frac{h^2}{2^{5/3}\pi k_B T} N_v(T)^{2/3} \tag{3}$$

with

$$N_v(T) = \int_{-\infty}^{E_{VBM}} g(E)\exp(E - E_{VBM}/k_B T)dE \tag{4}$$

where $h$ and $k_B$ are the Planck and Boltzmann constants, respectively, N$_v$($T$) is the so-called effective density of states for the valence band, and $g(E)$ is the actual density of states.

Analogous equations hold for the electron effective mass $m_e^*$. For the ideal case of a parabolic band, $m^*$ is independent of the temperature $T$ used for the DOS integration. Taking into account finite k-point sampling, we chose here $T = 900$ K to determine $m^*$. The effective masses are listed in Table I. Up to alloy compositions of $x = 0.125$, we observe a moderate increase of the effective masses that is fairly independent on the chemical element, as expected from the discussion above. At high concentrations of $x = 0.25$, the effective masses increase noticeably, indicating a more severe perturbation of the band structure. We conclude from the electronic structure analysis that the favorable carrier transport properties of SnO should be maintained for moderate alloy compositions $x < 0.25$ for all four elemental substitutions considered here.

Table I. The calculated DOS effective masses for electrons and holes, $m_e^*$ and $m_h^*$, respectively, for different $Sn_{1-x}M_xO$ alloy compositions. The effective masses are given in units of the electron rest mass $m_0$.

| $x$ | $m_e^*(m_0)$ | | | | $m_h^*(m_0)$ | | | |
|---|---|---|---|---|---|---|---|---|
| | Mg | Ca | Sr | Zn | Mg | Ca | Sr | Zn |
| 0 | | 0.4 | | | | 0.9 | | |
| 0.03125 | 0.5 | 0.5 | 0.4 | 0.5 | 1.2 | 1.2 | 1.2 | 1.2 |
| 0.0625 | 0.5 | 0.6 | 0.7 | 0.5 | 1.3 | 1.3 | 1.3 | 1.3 |
| 0.125 | 0.6 | 0.6 | 0.7 | 0.6 | 1.5 | 1.6 | 1.4 | 1.5 |
| 0.25 | 1.1 | 1.3 | 1.8 | 1.1 | 2.1 | 2.0 | 1.9 | 2.1 |

We complete the computational section of this work by addressing the question how the theoretical PV efficiency limit of SnO depends on the alloy composition. To this end, we calculated the so-called spectroscopically limited maximum efficiency (SLME) [38]. Similar like the well-known Shockley-Queisser (SQ) limit [39], the SLME defines a theoretical efficiency limit for solar cells. However, whereas the SQ limit assumes full absorption of photons above the band gap, the SLME uses the calculated absorption coefficients to determine the maximum efficiency for a film of a given thickness. Thus, for real materials and finite film thickness, the SLME is always smaller than the SQ limit. Assuming a film thickness of 2 μm, the SLME is shown as a function of the composition $x$ in Fig. 4. Since the band gaps and optical absorption spectra (and hence the SLME), are very similar for all the alloy systems, we only show the average value in Fig. 4. We find that even at the smallest alloying composition considered here, the efficiency increases rapidly from < 1 % to 11 %, mostly due to a much better overlap between the absorption and solar spectra. The SLME curve saturates beyond $x = 0.125$, suggesting that only fairly small compositions are needed to fully exploit the alloying effect for the potential of SnO as PV absorber. We also show the SQ limit for comparison, as well as the SLME for the emerging PV material $Cu_2SnZnS_4$ (CZTS) [40] based on a GW calculation [41] We find that the theoretical SLME efficiency limit of SnO-based alloys is comparable to that of CZTS, in particular when considering that the real absorption coefficient should be somewhat larger than the calculated one, as discussed in the methods section.

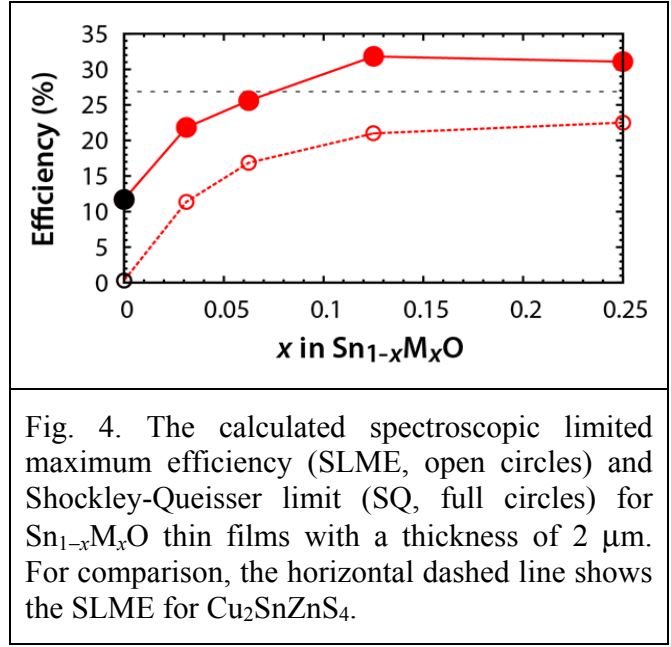

Fig. 4. The calculated spectroscopic limited maximum efficiency (SLME, open circles) and Shockley-Queisser limit (SQ, full circles) for $Sn_{1-x}M_xO$ thin films with a thickness of 2 μm. For comparison, the horizontal dashed line shows the SLME for $Cu_2SnZnS_4$.

**Experimental**

In order to provide an initial assessment as to whether the predicted alloys can realistically be synthesized, and to validate the predicted changes in the optical properties, we have grown $Sn_{1-x}Zn_xO$ films at the composition $x = 0.09 \pm 0.01$ and pure SnO reference films using RF sputtering. Figure 5a compares the experimentally measured XRD patterns with the simulated patterns from the computational atomic structures for $x = 0.125$. For the simulated XRD, the lattice constants of the calculated structures were slightly reduced (by 1.4 %) so to account for the systematic overestimation in the GGA. From the comparison, we can clearly assign the measured peaks to the calculated tetragonal SnO phase, without significant formation of secondary phases. The 2D detector images reveal that the pure SnO film is preferentially oriented, which explains some deviations in the XRD peak intensities, notably the absence of the (002) reflection in pure SnO. The $Sn_{0.91}Zn_{0.09}O$ film is less textured and contains a small amount of Sn impurity (estimated at < 2%), suggesting that further optimization of the synthesis conditions will be needed for growth of completely phase-pure alloys. The predicted change of the lattice parameters upon Zn substitution is well below 1% at $x = 0.125$. Thus, the expected XRD peak shift due to the alloying lies within the possible range due to residual stress in the film, which can occur during sputtering deposition depending on the specific process parameters [42].

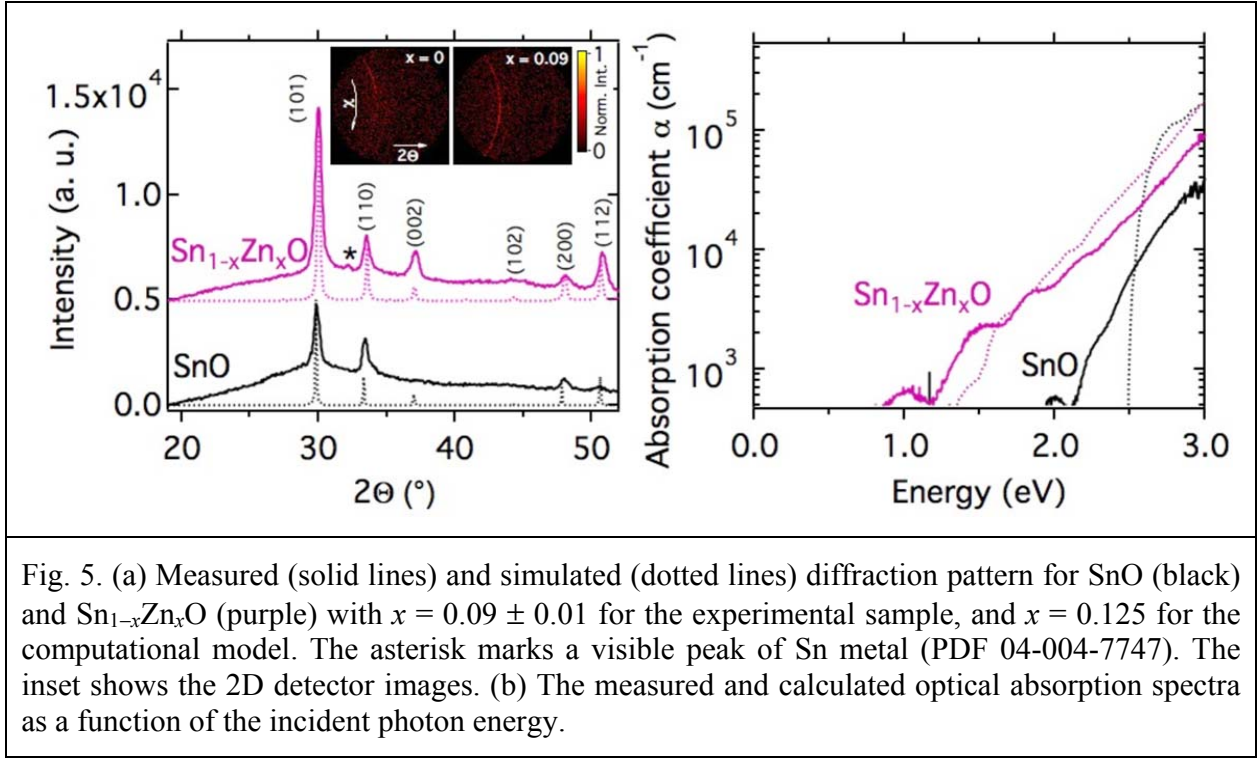

Fig. 5. (a) Measured (solid lines) and simulated (dotted lines) diffraction pattern for SnO (black) and $Sn_{1-x}Zn_xO$ (purple) with $x = 0.09 \pm 0.01$ for the experimental sample, and $x = 0.125$ for the computational model. The asterisk marks a visible peak of Sn metal (PDF 04-004-7747). The inset shows the 2D detector images. (b) The measured and calculated optical absorption spectra as a function of the incident photon energy.

A comparison of the calculated and measured absorption spectra is shown in Fig. 5b for both pure SnO and $Sn_{1-x}Zn_xO$. The computational spectrum for pure SnO shows a very steep absorption onset at the direct band gap of 2.5 eV obtained in the meta-GGA calculation. In the experimental spectrum the absorption onset is more gradual, likely due to the contribution of indirect absorption processes that are not included in the calculation. For the alloy, both the experimental and computational spectra show an overall increase of the absorption coefficient, and a decrease of the energy of the absorption onset, even though the fundamental band gap is expected to increase with the alloy composition $x$ based on the theoretical calculations (Fig. 1b). Thus, the initial synthesis and characterization of $Sn_{0.9}Zn_{0.1}O$ films demonstrate the feasibility of thin-film growth of the desired alloy compositions, and confirm the predicted trends in the optical spectra. It should be noted, however that further experimental work is needed to separate or eliminate the possibility of increased light scattering due to metallic Sn inclusions, which could affect the experimental optical spectra. Also, further work to experimentally characterize the structural details of Zn incorporation in the SnO films is highly desirable, so to validate the underlying structural models assumed for the alloy calculations.

## Conclusions

Compounds of low valent cations with occupied $s$-orbitals ($Sn^{2+}$, $Pb^{2+}$, $Sb^{3+}$, and $Bi^{3+}$) are recently attracting interest for their expected defect tolerance and due to the success of lead-halide perovskite photovoltaics that developed over the past years at an astounding pace. Among these materials, SnO is an oxide with exceptional electronic properties. However, hindering potential application in all-oxide PV is the strongly indirect nature of the band gap. In a computationally-driven materials exploration study reported here, we performed first-principles calculations for SnO based alloys formed when $Sn^{2+}$ is substituted by the divalent cations Mg, Ca, Sr, and Zn. We found that alloying at moderate compositions around 10% sufficiently increases the fundamental band gap and decreases the absorption threshold to enable a theoretical PV efficiency above 20%. Initial experimental results for $Sn_{1-x}Zn_xO$ alloys show that thin-film synthesis of single-phase alloys is feasible and tentatively confirm the predicted change of the optical properties. As a broader implication, this work highlights the particular suitability of defect tolerant materials for band gap engineering approaches using elemental substitutions, for both solar cells and other energy conversion technologies.

## Methods

### Computational

We have considered alloying SnO (litharge structure, space group 129) with Mg, Ca, Sr, and Zn, with compositions of 3.125%, 6.25%, 12.5%, and 25%. To model the alloy structures, we employed the special quasi-random structures (SQS) [43, 44] as generated with the *mcsqs* utility as implemented in the ATAT [45]. Relatively large 256-atom SQS supercells were chosen such to include low alloy compositions. In order to facilitate the Brillouin zone sampling in the density functional theory (DFT) calculations, we have chosen the supercells in a shape close to a cube, and search for SQS structures such that the pair correlation functions agree with the ideal random alloy up to the 8th nearest neighbor in the cation sublattice for all compositions considered. To further take into account the nature of random alloying, we sample the results over four different SQS for each composition.

The first-principles calculations were performed the projector augmented wave (PAW) method [46] as implemented in the VASP code [47, 48]. Using a 2×2×2 k-mesh for Brillouin zone integration, the SQS were fully relaxed within the generalized gradient approximation (GGA) [49], using an on-site Coulomb correction [50] with $U = 6$ eV for the Zn-$d$ states. In order to obtain an approximately band gap corrected electronic structure and optical properties, we utilized the modified Becke-Johnson local density approximation (mBJ-LDA) [33, 51, 52], a potential-based meta-GGA functional that gives band gaps close to many-body approaches, but at a computational cost comparable to standard DFT calculations. The optical properties were calculated with a 3×3×3 k-mesh using the linear optics implementation of Ref. [53]. Note that the calculated absorption spectrum is based only on direct transitions in the independent particle approximation, and does not include phonon-assisted indirect absorption or excitonic effects, which can further enhance the absorption coefficients [53]. The $E(k)$ relationship was determined for two alloy supercells by unfolding of the band-structure using the BandUP code [54].

### Experimental

In order to test the feasibility of isovalent SnO alloys, and to verify our theoretical predictions, we synthesized experimentally SnO (1.5 μm thick) and $Sn_{1-x}Zn_xO$ (0.75 μm thick) thin-films, using radio frequency (RF) reactive magnetron sputtering (13.56 MHz) from metallic targets in an argon/oxygen atmosphere, similar to the previously reported studies of other novel absorber and contact materials [55, 56]. The films were grown on Corning Eagle XG glass substrates at a temperature of 260 °C. During deposition, the total pressure was kept at 1.6 Pa with a mass flow ratio $fO_2/(fO_2+fAr)$ of 1.75% for the $Sn_{1-x}Zn_xO$ films and 2.25% for the pure

SnO film, respectively. It is necessary to reduce the oxygen partial pressure in the sputtering atmosphere, to suppress the change of Sn valence state from $Sn^0$ via $Sn^{2+}$ to $Sn^{4+}$ due to the addition of Zn during the SnO synthesis. By simultaneously adjusting the Zn and the O content, we were able to prepare $Sn_{1-x}Zn_xO$ films with $x$ = 9 % in the tetragonal SnO-like structure without a significant formation of secondary phases according to X-ray diffraction (XRD). The XRD in the Bragg-Brentano geometry with a Bruker D8 using Cu K$\alpha$ radiation and a 2-D detector was used to characterize the crystalline phase of the films. Optical absorption spectra were measured in the UV-NIR (300-1600 nm) range with an Ocean Optics spectrometer and the absorption coefficients ($\alpha$) were calculated from the measured transmittance (T) and reflectance (R) according to $\alpha = -(1/d)\ln[T/(1-R)]$ where $d$ is the film thickness. More details on characterization techniques have been previously reported [57, 58].

## Acknowledgements

This work is supported by the U.S. Department of Energy, Office of Energy Efficiency and Renewable Energy, under Contract No. DE-AC36-08GO28308 to the National Renewable Energy Laboratory (NREL). This work used computational resources sponsored by the Department of Energy's Office of Energy Efficiency and Renewable Energy, located at NREL.